\documentclass[a4paper,fleqn]{cas-dc}

\usepackage[T1]{fontenc}
\usepackage[utf8]{inputenc}

\usepackage[authoryear]{natbib}

\usepackage[nameinlink]{cleveref}
\usepackage[official]{eurosym}
\usepackage{pgfplots}
\usepgfplotslibrary{groupplots}
\usepackage{xcolor}
\usepackage{placeins}
\usepackage[english]{babel}

\RequirePackage[normalem]{ulem} 
\RequirePackage{color}\definecolor{RED}{rgb}{1,0,0}\definecolor{BLUE}{rgb}{0,0,1} 
\RequirePackage{listings} 
\RequirePackage{color} 
\lstdefinelanguage{DIFcode}{ 
  moredelim=[il][\color{red}\sout]{\%DIF\ <\ }, 
  moredelim=[il][\color{blue}\uwave]{\%DIF\ >\ } 
} 
\lstdefinestyle{DIFverbatimstyle}{ 
	language=DIFcode, 
	basicstyle=\ttfamily, 
	columns=fullflexible, 
	keepspaces=true 
} 
\lstnewenvironment{DIFverbatim}{\lstset{style=DIFverbatimstyle}}{} 
\lstnewenvironment{DIFverbatim*}{\lstset{style=DIFverbatimstyle,showspaces=true}}{} 

\definecolor{dunkelgruen}{RGB}{0,120,107}
\definecolor{tuerkis}{RGB}{111,200,182}

\definecolor{batterycolor}{RGB}{218,216,255}
\definecolor{electrolysiscolor}{RGB}{148,241,255}
\definecolor{hydrogencolor}{RGB}{0,102,126}
\definecolor{pvcolor}{RGB}{255,235,59}
\definecolor{windoffcolor}{RGB}{16,76,90}

\begin{document}
\let\WriteBookmarks\relax
\def\floatpagepagefraction{1}
\def\textpagefraction{.001}
\shorttitle{Centralized and decentral approaches to succeed the low-carbon Energiewende in Germany in the European context}
\shortauthors{Kendziorski et al.}

\newcommand{\integriert}{\textit{integrated }}
\newcommand{\desintegriert}{\textit{disintegrated }}

\title [mode = title]{Centralized and decentral approaches to succeed the 100\% energiewende in Germany in the European context –
A model-based analysis of generation, network, and storage investments
}

\address[1]{German Institute for Economic Research (DIW Berlin), Mohrenstraße 58, 10117 Berlin, Germany}

\address[2]{Leuphana University, Universitätsallee 1, 21335 Lüneburg, Germany}

\address[3]{Berlin University of Technology, Workgroup for Infrastructure Policy (WIP), Straße des 17. Juni 135, 10623 Berlin, Germany}

\author[1,3]{Mario Kendziorski}[orcid=0000-0001-5508-2953]
\cormark[1]
\ead{mak@wip.tu-berlin.de}

\author[1,3]{Leonard Göke}[orcid=0000-0002-3219-7587]
\ead{lgo@wip.tu-berlin.de}

\author[1,3]{Christian {von Hirschhausen}}[orcid=0000-0002-0814-8654]
\ead{cvh@wip.tu-berlin.de}

\author[1,2]{Claudia Kemfert}[orcid=0000-0002-6742-0478]
\ead{ckemfert@diw.de}

\author[3]{Elmar Zozmann}
\ead{ez@wip.tu-berlin.de}

\cortext[cor1]{Corresponding author}                   



\begin{abstract}
In this paper, we explore centralized and more decentral approaches to succeed the energiewende in Germany, in the European context. We use the AnyMOD framework to model a future renewable-based European energy system, based on a techno-economic optimization, i.e. cost minimization with given demand, including both investment and the subsequent dispatch of capacity. The model includes 29 regions for European countries, and 38 NUTS-2 regions in Germany. First the entire energy system on the European level is optimized. Based on these results, the electricity system for the German regions is optimized to achieve great regional detail to analyse spatial effects. The model allows a comparison between a stylized central scenario with high amounts of wind offshore deployed, and a decentral scenario using mainly the existing grid, and thus relying more on local capacities. The results reveal that the cost for the second optimization of these two scenarios are about the same: The central scenario is characterized by network expansion in order to transport the electricity from the wind offshore sites, whereas the decentral scenario leads to more photovoltaic and battery deployment closer to the areas with a high demand for energy. A scenarios with higher energy efficiency and lower demand projections lead to a significant reduction of investment requirements, and to different localizations thereof.
\end{abstract}

\maketitle

\section{Introduction}
In the context of the European Green Deal and the move towards climate neutrality, attempts are intensifying to deploy renewable energies at all levels, from highly centralized generation, such as large-scale offshore wind, to decentralized generation, e.g. local wind parks and solar rooftop installations. While central approaches favour large-scale producers and use the extension of the network to reduce congestions, decentral approaches are driven largely by residents and incentivize generation close to the consumption. Amongst other countries, Germany is exploring different pathways for an acceleration of the energiewende, the low-carbon energy transformation with a high level of citizen engagement. This implies the involvment of more decentral units, such as cities, communities, and city neighborhoods. Nuclear power and CO2-capture are considered to be no options and are not part of the energy mix due to their high costs.

The German low-carbon transformation, called "energie\-wende", was kicked off in the 1970s by citizen engagement and attempts to complement centralized energy production by local generation, closer to demand, and is largely renewables-based \citep{morris_energy_2016, von_hirschhausen_energiewende_2018}. Likewise, the EU has adopted a strategy to spur citizen engagement, and give local communities the option to self-produce and, to a certain extent, self-consume \citep{ec_directive_2019}. In this context, questions arise from the comparison of a more central and a more decentral approach: For example, how does the generation mix change, if one choses a more decentral approach? How costly would  a 100\% renewable energy supply  be if network costs and congestion were taken into account (which is under today's regulatory practices not the case)? In addition, how would these costs compare with a more centralized approach? \par
Modeling a fully renewable energy supply for Germany is not a new topic by itself, and is addressed, amongst others by \citep{sru_pathways_2011,jacobson_100_2017, bartholdsen_pathways_2019, zerrahn_economics_2018}. Also, different studies have highlighted different dimensions of prosumage, such as a rather small size of installations, a rather high degree of self-con\-sump\-tion, and the proximity to demand \citep{schill_prosumage_2017}. Previous research also shows the interrelation between network pricing and the structure of electricity generation: the more networks cost, and the more regional price signals are taken into account, the less need for network extension occurs \citep{neuhoff_renewable_2013, kemfert_welfare_2016}; this also holds for flexibility on the supply and the demand side, including curtailment of generation \citep{grimm_transmission_2016, grimm_reduction_2016}. Last but not least, network regulation and market design also affect issues of distributional justice, and can contribute to allocation schemes that are perceived as more fair than others \citep{drechsler_efficient_2017}. However, the interdependence between generation, network structure, storage and demand has not yet been modeled at such a decentral level, which is the major contribution of this paper. \par

The analysis in this paper relies on the energy system model AnyMOD that calculates an optimal mix of energy, storage, and network infrastructure for a given energy demand (\Cref{section2}). The model decides on how to satisfy an exogenous demand by deploying various technologies that generate, convert, transport, and store various energy carriers. The model covers the entire European energy system and has been extended, for the purpose of this paper, to include 38 regions within Germany (called “NUTS2-zones”, Nomenclature of Territorial Units for Statistics). Another novelty is the coverage of full sector coupling: AnyMOD combines the traditional electricity sector with demands from industry, buildings, and mobility, as well as exogenously defined demand for synthetic gases, e.g. for use in industrial processes. We use two quite different demand scenarios to cover a variety of possible futures: the reference scenario (called “REF”) includes overall energy demand of 1,209 TWh, of which 1,070 TWh are electricity, and 139 TWh is demand for hydrogen; an alternative demand scenario includes significant energy savings and energy efficiency (called “EFF”) and, thus, only 610 TWh of final demand. Results from another project, which calculates a path towards a fully renewable-based European energy system, are the foundation for the model. The parameters are calibrated to the year 2040. \par

Sections \ref{section3} and \ref{section4} include the main questions and results of the modelling:  \Cref{section3} compares a scenario in which network costs are completely ignored (corresponding to the status quo in Germany), with a situation in which network investment and operational costs are taken into account in the energy system planning process. Model results suggest quite a strong effect of the latter modification on the nature of energy generation and storage, and a push for more decentral activity: Central generation, mainly offshore wind, is reduced from 50 GW to below 20 GW, whereas onshore wind, PV, and battery storage, respectively, increase; network extension is reduced by over 50\%. \Cref{sectioneff} focuses on a particularly interesting result, i.e. the effect of energy demand on generation, storage, and transmission infrastructure. The efficiency demand scenario (EFF), in which demand is significantly lower than the reference scenario, leads to a drastic reduction of the need for PV, onshore wind, as well as electrolyzers with reductions of over 50\% compared to the reference scenarios. Offshore wind, battery storage, and network extensions almost disappear under this scenario. \par

Last but not least, in Section \ref{section4} we present and discuss model results on the costs of more centralized and more decentral generation scenarios (taking into account network costs in both cases). The stylized decentral scenario is defined by very high levels of regional supply-de\-mand balancing, with the transmission infrastructure fixed at the 2022 level; by contrast, the central scenario includes (exogenously defined) 50 GW offshore wind and no constraint on network expansion. Somewhat logically, the decentral scenario leads to higher investments in PV and battery storage, whereas electrolyzers and hydrogen turbines are only marginally affected. Most interestingly, both scenarios have very similar total costs for the German electricity system. \Cref{section6} concludes and draws policy implications.

\section{Methodology and data}
\label{section2}
We use the AnyMOD framework \citep{goke_anymod_2020} to create an energy system model which, apart from the electricity sector, includes demand from space heating, industry heat, electric vehicles, and hydrogen. The model optimizes the investment in generating technologies (e.g. wind turbines or solar PV), storage technologies (e.g. batteries), converting technologies (e.g. electrolyser, hydrogen turbines), and transmission infrastructure. The existing installed capacities are based on \citet{hainsch_make_2020} which displays a decarbonization path of the energy system until 2040.

\subsection{Spatial and temporal dimension}
\label{section2:dimension}

\begin{figure}
    \centering
		\includegraphics[width=\columnwidth]{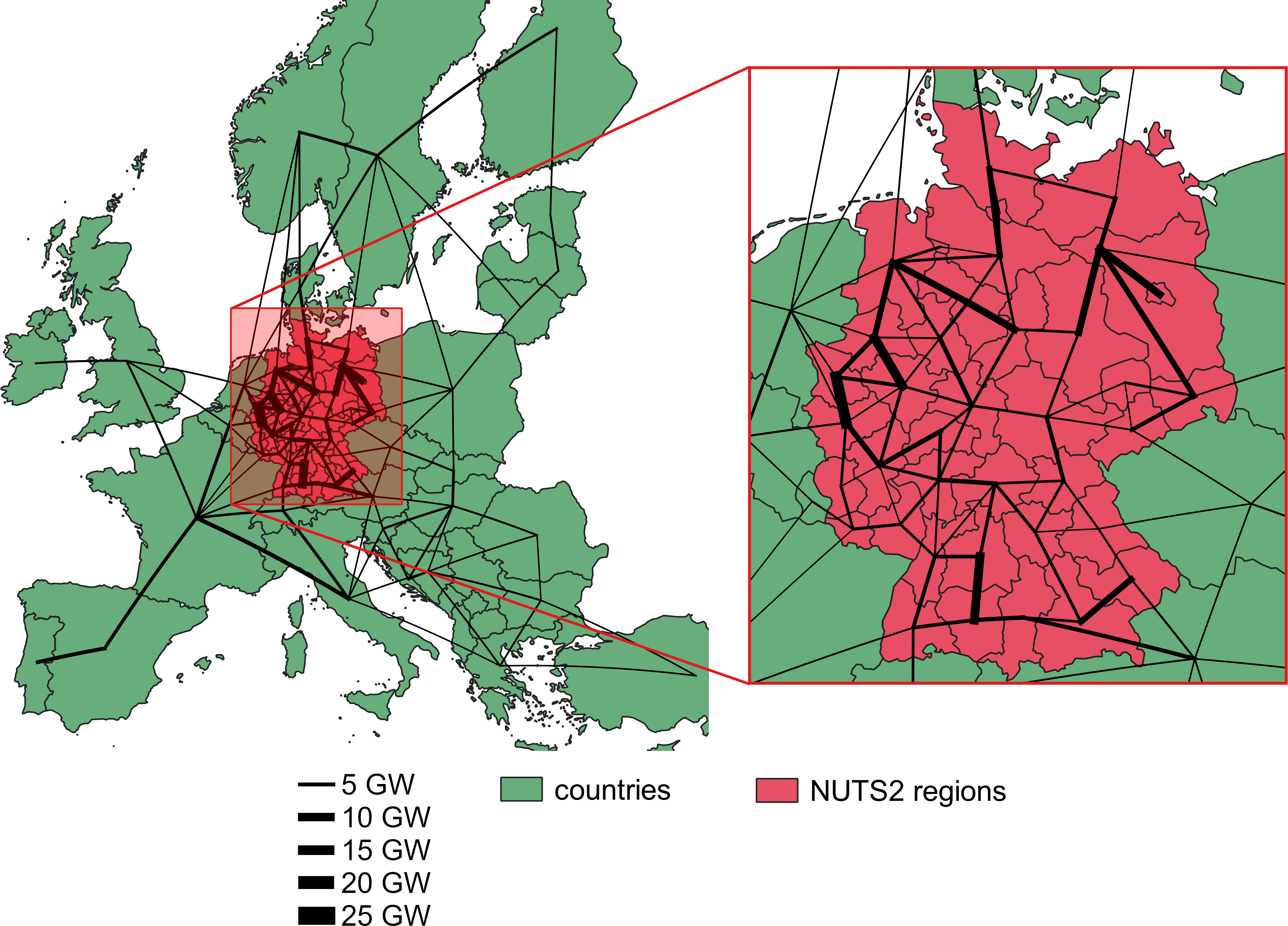}
	\caption{Spatial resolution of the model with existing trading capacities.}
	\label{fig:overview_regions}
\end{figure}

The temporal resolution of the model varies for the different  energy carriers. The electricity sector is modelled in an hourly resolution in order to capture the intermittent feed-in of wind and solar based technologies and the resulting need for flexibility options. Electrical vehicles, space heat, and industry demand are balanced in four hour blocks which reflects the inherent flexibility of those sectors while it reduces computational effort. The gaseous carriers (hydrogen and synthetic methane) are modelled on a daily basis. \par

We solve the model in a two-step process: First, the model is solved at the European level, where each country is represented by a node. In this run the entire energy system of each European country is optimized in a greenfield setting. Second, based on these results investment decision of the European countries besides Germany are fixed. Subsequently, Germany is represented by 38 regions. A deviation in the investment decision compared to the first run is possible while only the dispatch decision can still be optimized in the other countries. This allows for a high spatial resolution while the neighboring countries and their exchange potential is still taken into account. \par

The trading potential between the European countries is limited to the projects in the Ten-Year-Network-Development-Plan (TYNDP) \citep{entso-e_tyndp_2019} which will be finished by 2025. An abstract representation of the existing transmission infrastructure between the 38 German regions is implemented (Fig. \ref{fig:overview_regions}) and can be extended depending on the analysed scenario. However, in the model the transmission network does not follow electrical alternating current (AC) load flow rules and is therefore a simplified representation. To account for this overestimation of transport capacity, it was reduced to 70\% of its original value. Synthetic methane can be traded using the existing pipeline infrastructure. Additional investments are necessary to upgrade the pipelines to enable the transport of pure hydrogen. Imports of hydrogen or electricity from outside the EU are not considered since we model an energy-independent Europe. However, if cheap hydrogen imports were available in the future, this would simplify the challenge of decarbonizing the system.

\subsection{Technologies and cost assumptions}

\begin{table*}
\caption{Cost assumption of the most important technologies in 2040.}
\label{tab:cost_assumtions}
\centering
\begin{tabular}{llllll} 
\toprule
Technology       & Overnight cost  & Overnight cost & Operational cost  & Lifetime & Source      \\
 & [\euro{}/kW] & [\euro{}/kWh] & [\euro{}/kW/y] & [y] & \\
\midrule
PV (rooftop)     & 588            &                & 8.14             & 25       & \cite{fraunhofer_ise_stromgehstehungskosten_2018}  \\
PV (open space)  & 317            &                & 6.34             & 25       & \cite{fraunhofer_ise_stromgehstehungskosten_2018}  \\
Wind onshore     & 1140           &                & 44.4             & 25       & \cite{fraunhofer_ise_stromgehstehungskosten_2018}   \\
Wind offshore\footnotemark   & 2335 - 3540    &                & 46.7 - 70.8      & 30       & \cite{fraunhofer_ise_stromgehstehungskosten_2018}   \\
Hydrogen turbine & 185            &                & 3.3              & 30       & \cite{auer_quantitative_2020}   \\
Electrolyser     & 418            &                & 14.6             & 30       & \cite{osmose_flexibility_2019}      \\
Battery          & 74.7           & 164.1          & 1.1              & 18       & \cite{osmose_flexibility_2019}      \\
Hydrogen storage & 4.9            & 0.00497        &                  &          & \cite{auer_quantitative_2020}   \\
\bottomrule
\end{tabular}

\end{table*}
\footnotetext{The cost are depending of the distance to the shore and the depth of the sea.}

 To cover the given demands for electricity, space and industry heat, methane, and hydrogen, a vast array of technologies can be used in the energy system model. Table \ref{tab:cost_assumtions} contains the most important technologies and the corresponding cost data. Other technologies (e.g. fermenter, methanation, compressed-air energy storage, or pumped hydro storages)  are considered but play a minor role in the results. Capital costs of investments are annualised with an interest rate of 2\%.
 
The availability factors for intermittent renewables are taken from \citet{auer_quantitative_2020} for Europe and from \citet{ninja1} and \citet{ninja2} for the NUTS2-Regions in Germany.

We base our estimation of standard grid costs on the \citep{moles_energy_2014}. Here the costs of a 500 MVA line are given as 480,000 €/km, which translates into 2.74 Mil. €/km/GW assuming a line capacity of 175 MW. These costs are then multiplied with the respective length of each line in the stylized grid to obtain specific investment costs. The lifetime used to annualize these costs amounts to 60 years. We added the corresponding information to the paper.

\subsection{Renewable potentials}
\label{section2:res}

One of the key input parameters to an energy system model are the upper bounds of renewable energies. Table \ref{tab:res_potentails} compiles a selection those potential estimates for Germany from various sources. For this study, the potentials are 226 GW for PV open space, 900 GW PV rooftop, 223 GW wind onshore, and 80 GW wind offshore. From 50 GW onwards, wake effects for wind offshore capacities are considered which result in lower full load hours \citep{agora_energiewende_making_2020}.

\begin{table}[width=.9\linewidth,pos=h]
\caption{Renewable potentials for Germany from different sources.}
\label{tab:res_potentails}
\centering
\begin{tabular}{llll} 
\toprule
Technology                       & Installed &  Source        \\
 & capacity &  \\
 & [GW] & & \\
\midrule
                                 & 100                        & \cite{bodis_high-resolution_2019}         \\
PV                                 & 208                        & \cite{mainzer_high-resolution_2014}       \\
rooftop                          & 161                        & \cite{lodl_abschatzung_2010}          \\
                                 & 900                          & \cite{fraunhofer_ise_integrierte_2021}    \\
\midrule
PV                        & 226                         & \cite{fraunhofer_ise_integrierte_2021}    \\
open space &  & \\
\midrule
PV                         & 1444                      & \cite{breyer_100_2020}  \\
 total                                & 312                        & \cite{auer_quantitative_2020}  \\
\midrule
                               & 240                     & \cite{breyer_100_2020} \\
Wind                                 & 223                       & \cite{auer_quantitative_2020}  \\
onshore                                 & 396                     & \cite{masurowski_spatially_2016} \\
                                 & 585                      & \cite{masurowski_spatially_2016} \\
\bottomrule
\end{tabular}

\end{table}

The national potentials for PV (rooftop and open space) and wind onshore are distributed to each of the 38 regions using the land use as key parameters from the \citet{Corine}. The available shares of the categories urban, sub-urban, agricultural, and forested areas are based on \citet{Nahmacher2014} and \citet{Bodis2019}. After determining total available land in each region, the quality of the sites is assessed using geological data \citep{solar,wind}. The result is clustered into different groups and the corresponding time series are scaled according to the site quality in such as way that the total energy potential of each region is unchanged.

Since the area for wind onshore and open space PV is limited to agricultural and forested areas, the potential is high in areas with a lower population density, In contrast, areas with a higher population density have a high potential for rooftop PV. The wind offshore capacities are allocated to the closest onshore region.

\subsection{Demand}
\label{section2:demand}

As a starting point we use the scenario from \citet{hainsch_make_2020} which provides a fully renewable energy system for Europe. Key decisions such as the technology choices for space heating, industry heat, private and freight transport were computed with the established energy system model GENeSYS-MOD \citep{burandt_genesys-mod_2018} and data from \citet{auer_quantitative_2020}. A large share of the space heat is covered by heat pumps which results in additional electricity demand (91 TWh in Germany) for our model run. Additional electricity demand comes from the industry sector (456 TWh, e.g. electric furnaces etc.), private and freight transport (223 TWh, electric vehicles and cargo by rail). A minor share of the demand from these sectors is covered by hydrogen (134 TWh) or synthetic methane (6 TWh). The conventional electricity demand accounts for another 300 TWh.

This demand is distributed to the 38 regions using the population and GDP as indicators as industry and GDP are heavily correlated in Germany \citep{kunz_electricity_2017}. An hourly demand is then generated based on standardized load profiles  of households, commerce, and industry in Germany \citep{bdew_standardlastprofil_2020}.

We assume that the demand from heat pumps, industry, and electric vehicles can be satisfied within 4-hour blocks. This reflects the demand side response and thus the flexibility from integrating energy sectors.

\section{Results I: Generation and storages under different network regulation}
\label{section3}

The decarbonization envisaged in the context of the sustainability goals not only implies a restructuring of the energy supply in terms of the technologies used, but also in terms of its spatial structure. Unlike conventional generators, the costs and potentials of wind and solar plants depend strongly on their locations. Consequently, when placing renewable plants and flexibility options such as batteries, electrolyzers, or hydrogen turbines, it is important to incorporate the grid expansion costs in the assessment. Thus, taking into account network costs will in general lead to a more decentral structure of generation. On the other hand, a regime that ignores network costs in the dispatch decision, favours central generation structures \citep{kemfert_welfare_2016}. While these effects have been identified in the literature, scenario-based quantification including sector coupling have not.

The current centralized approach favours sites where the yields are highest. Decentralized approaches, on the other hand, aim to generate electrical energy where demand is high in order to harness the added value of these plants locally and not leave power generation exclusively to large energy operators. This approach not only leads to a reduction in grid expansion costs, but it can also increases the acceptance of the energiewende if local citizens have the possibility to participate and are able to profit from it. Moving generators closer to consumption can reduce the amount of transmission capacity needed, resulting in fewer bottlenecks in the transmission system. Consequently, the need for grid expansion is lower compared to an expansion path in which proximity to consumption is not a criterion. However, it is not yet possible to estimate how large the differences in this regard are between a 100\% renewable based scenario that takes this into account and a  scenario in which proximity to consumption is irrelevant. Similarly, it remains unclear what the differences between the two scenarios are in terms of total costs (energy production costs, storage and load shifting costs, redispatch and curtailment costs, and grid costs). This question is addressed in the following.

\subsection{Scenarios}
\label{section3:scenarios}

The analysis of the relationship between network planning, spatial structure of generation, flexibility options and decentralization is done in two steps: First, two grid planning scenarios are contrasted, both using the existing grid as a starting point. The decision about the generation and grid infrastructure expansion is considered once by neglecting the grid costs, as it corresponds to the common practice in Germany when placing renewable generators ("disintegrated" scenario) and once by considering the infrastructure costs ("integrated" scenario):

\begin{itemize}
    \item In the first case, the generation technologies as well as storage technologies are placed under the assumption that there are no grid bottlenecks (Germany is one single price zone), which corresponds to today's situation. As a result, plants are placed in such a way that they generate the highest possible yield, but also lead to increased grid expansion requirements that are not factored in (in the following, this variant is called "disintegrated"). This procedure roughly represents the current planning process of the grid development plan. A large part of the grid expansion costs are passed on to the end consumers as a grid fee, which results in a redistribution in favour of the power plant operators.
    
    \item The second variant considers the investment in grid expansion as well as generation and storage capacities combined (called "integrated"). This considers the trade-off between the highest return and the necessary grid expansion costs. In such a procedure, the spatial component also plays a role, as a location close to consumption can reduce additional grid expansion. In this setting, investments into storage capacities can also be used as an alternative to upgrading line capacity. This is likely to result in a more efficient system as storages can act as a temporal or spatial buffer while additional line capacity is only capable of doing the latter.
    
\end{itemize}

\subsection{Incorporating network expansion costs leads to more decentral structures}

\begin{figure}
    \centering
\begin{tikzpicture}[font=\scriptsize]

  \draw[gray,thin] (5.3,0.41) -- (5.3,5.15);

  \begin{axis}[
    ybar,
    symbolic x coords={Battery, Electrolyser, PV, Hydrogen turbine, Wind onshore, Wind offshore, Network expansion},
    bar width=0.25cm,
    xtick=data,
    x tick label style={rotate=45,anchor=north east},
    axis y line*=left,
    ylabel=GW,
    ylabel near ticks,
    ymajorgrids=true,
    grid style=dashed,
    tickwidth         = 0pt,
    legend style={
			draw=none,
			at={(0.5,1.12)},
			anchor=north},
	legend columns=-1,
	width=0.45\textwidth
    ]
          \addplot[draw=none,fill=dunkelgruen] coordinates {
    (Battery,14.8)
    (Electrolyser,80.5)
    (PV,269.8) 
    (Hydrogen turbine,113.1)
    (Wind onshore,198.0) 
    (Wind offshore,50.0)
    (Network expansion, 0)
    };
    
    \addplot[draw=none,fill=tuerkis] coordinates {
        (Battery,26.7)
        (Electrolyser,82.8)
        (PV,305.9) 
        (Hydrogen turbine,92.3)
        (Wind onshore,217.9)  
        (Wind offshore,18.0)
        (Network expansion, 0)
      };

    \legend{Desintegrated, Integrated}
  \end{axis}
  
  \pgfplotsset{every axis y label/.append style={rotate=180,yshift=8cm}}
  
  \begin{axis}[
    ybar,
    symbolic x coords={Battery, Electrolyser, PV, Hydrogen turbine, Wind onshore, Wind offshore, Network expansion},
    bar width=0.25cm,
    hide x axis,
    x tick label style={rotate=45,anchor=north east},
    axis y line*=right,
    ylabel=TWkm,
    tickwidth         = 0pt,
    width=0.45\textwidth,
    ymax=12.5
    ]
    
         \addplot[draw=none,fill=dunkelgruen] coordinates {
        (Battery,0)
        (Electrolyser,0)
        (PV,0) 
        (Hydrogen turbine,0)
        (Wind onshore,0) 
        (Wind offshore,0)
        (Network expansion,11.6) 
        };
        
    \addplot[draw=none,fill=tuerkis] coordinates {
        (Battery,0)
        (Electrolyser,0)
        (PV,0) 
        (Hydrogen turbine,0)
        (Wind onshore,0) 
        (Wind offshore,0)
        (Network expansion, 5.0) 
      };
  \end{axis}

\end{tikzpicture}
	\caption{Comparison of installed capacities and network expansion between the \desintegriert and the \integriert scenario.}
	\label{fig:cap_plot1}
\end{figure}
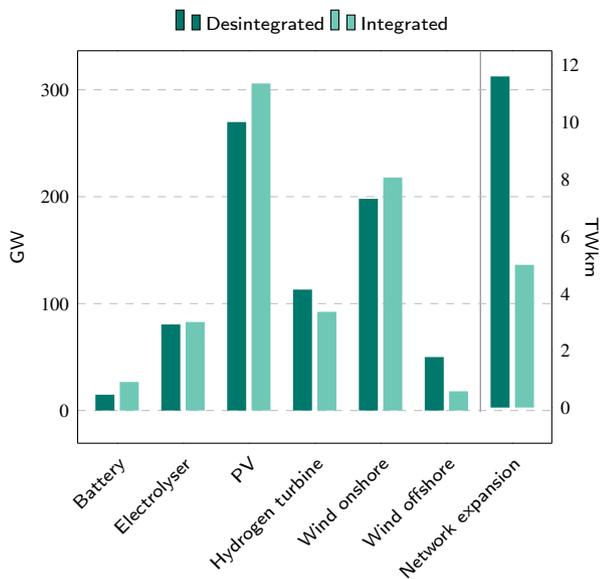

By taking grid expansion costs into account, decentralized generation structures that are close to the load are strengthened. Figure \ref{fig:cap_plot1} provides an overview of the alternative planning approaches: In both variants of grid planning a combination of solar PV and wind onshore constitutes the largest share of generation technologies. The deployment of wind offshore decreases in the \textit{Integrated} scenario from 50 GW to around 15 GW. This is mainly compensated by an increase of installed capacity from solar PV, a moderate increase of wind onshore, and a slight increase in batteries. Since wind offshore facilities are geographically located in the north, transmission infrastructure is necessary to transport the generation to the demand. As a result, higher capacities of wind offshore force more investments in the network expansion, meaning more than doubling the amount compared to the \textit{Integrated} scenario. The reduction in hydrogen turbines are caused by the additional investment in batteries, as these technologies both are used to cover the residual peak load in times with low intermittent renewable feed-in. The investment in electrolysers is unaffected since approximately the same amount of hydrogen is produced over the year. The additional peak generation from PV is largely absorbed by more battery storage or consumed by other flexiblity options from the demand side (EV, heat pumps, or industry).

\begin{figure}
    \centering
	\includegraphics[width=.46\linewidth]{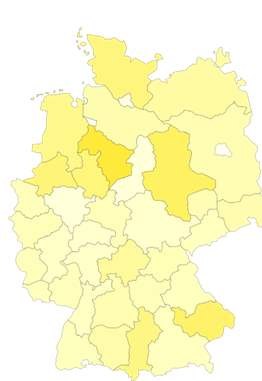}
	\includegraphics[width=.51\linewidth]{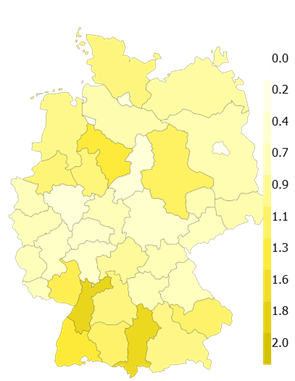}
	\caption{Solar PV capacities in MW/km² (left \desintegriert, right \integriert).}
	\label{fig:map_pv_res1}
\end{figure}

While wind offshore parks are strongly restricted to certain areas, solar PV and wind onshore capacities can be deployed almost everywhere. Especially rooftop solar PV makes use of already sealed areas. This comes at the cost of higher installation costs and a sub-optimal orientation of the PV modules. Yet local proximity between generation and demand can be beneficial: This can be observed in Fig. \ref{fig:map_pv_res1}, in the \textit{Integrated} scenario: Most of the additional solar PV capacities are located in the South, a region with high demand caused by many industrial sites and a rather long distance to the shore. While generation cost might be lower for wind offshore parks, the required network reinforcements to transport that energy down south causes higher costs overall. In this case, it is optimal to build more solar PV capacity even though sites with high full load hours are already taken.

\begin{figure}
    \centering
	\includegraphics[width=.46\linewidth]{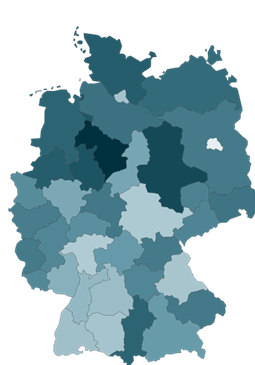}
	\includegraphics[width=.51\linewidth]{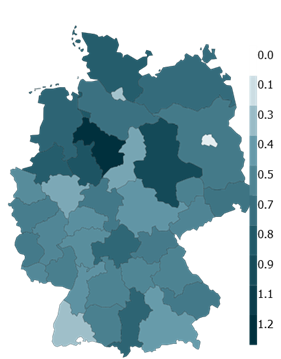}
	\caption{Wind onshore capacities in MW/km²(left \desintegriert, right \integriert).}
	\label{fig:map_wind_res1}
\end{figure}

Similarly, the deployment of wind onshore capacities in the south is higher in the \textit{Integrated} scenario (Fig. \ref{fig:map_wind_res1}). Even though full load hours of wind turbines are below average, more wind turbines are built the due to the proximity of demand. The wind onshore potentials in every region are almost fully utilised in the \textit{Integrated} scenario, showing the importance of wind onshore for a fully renewable energy system.

\begin{figure}
    \centering
	\includegraphics[width=\columnwidth]{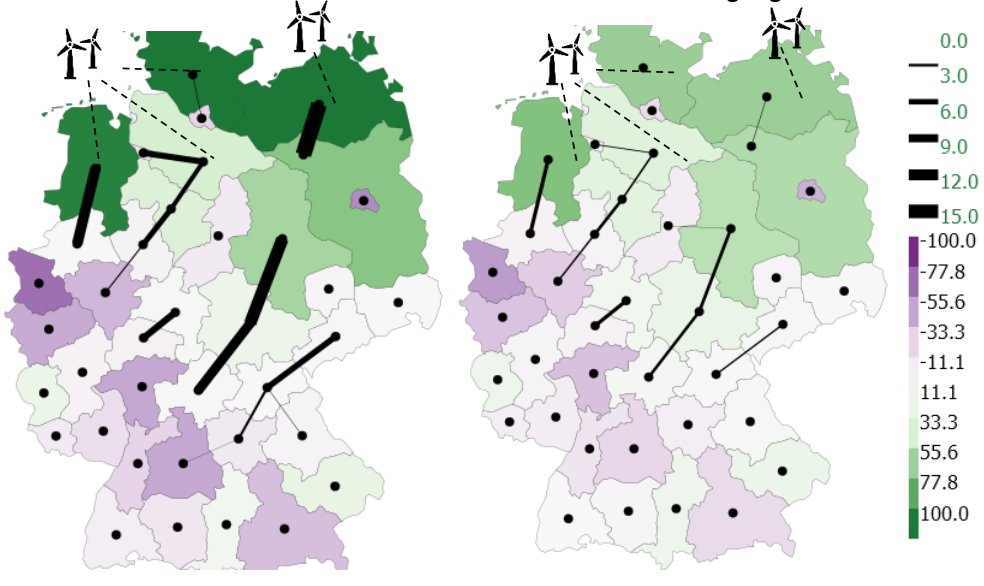}
	\caption{Yearly net exchange for each region in TWh and network reinforcements in GW (left \desintegriert, right \integriert).}
	\label{fig:map_exchange_res1}
\end{figure}

The more evenly distributed deployment of renewable energies reduces the necessary network expansion and results in more balanced exchange between the regions (Fig. \ref{fig:map_exchange_res1}). In the \desintegriert scenario a surplus is generated in the north and needs to be transported south. 

\subsection{Efficiency measures heavily reduce infrastructure investments}
\label{sectioneff}

\begin{figure}
    \centering
\begin{tikzpicture}[font=\scriptsize]

  \draw[gray,thin] (5.3,0.41) -- (5.3,5.15);

  \begin{axis}[
    ybar,
    symbolic x coords={Battery, Electrolyser, PV, Hydrogen turbine, Wind onshore, Wind offshore, Network expansion},
    bar width=0.25cm,
    xtick=data,
    x tick label style={rotate=45,anchor=north east},
    axis y line*=left,
    ylabel=GW,
    ylabel near ticks,
    ymajorgrids=true,
    grid style=dashed,
    tickwidth         = 0pt,
    legend style={
			draw=none,
			at={(0.5,1.12)},
			anchor=north},
	legend columns=-1,
	width=0.45\textwidth
    ]
          \addplot[draw=none,fill=dunkelgruen] coordinates {
    (Battery,14.8)
    (Electrolyser,80.5)
    (PV,269.8) 
    (Hydrogen turbine,113.1)
    (Wind onshore,198.0) 
    (Wind offshore,50.0)
    (Network expansion, 0)
    };
    
    \addplot[draw=none,fill=tuerkis] coordinates {
        (Battery,1.6)
        (Electrolyser,27.1)
        (PV,128.6) 
        (Hydrogen turbine,29.7)
        (Wind onshore,124.1)  
        (Wind offshore,7.7)
        (Network expansion, 0)
      };

    \legend{REF, EFF}
  \end{axis}
  
  \pgfplotsset{every axis y label/.append style={rotate=180,yshift=8cm}}
  
  \begin{axis}[
    ybar,
    symbolic x coords={Battery, Electrolyser, PV, Hydrogen turbine, Wind onshore, Wind offshore, Network expansion},
    bar width=0.25cm,
    hide x axis,
    x tick label style={rotate=45,anchor=north east},
    axis y line*=right,
    ylabel=TWkm,
    tickwidth         = 0pt,
    width=0.45\textwidth,
    ymax=12.5
    ]
    
         \addplot[draw=none,fill=dunkelgruen] coordinates {
        (Battery,0)
        (Electrolyser,0)
        (PV,0) 
        (Hydrogen turbine,0)
        (Wind onshore,0) 
        (Wind offshore,0)
        (Network expansion,5.0) 
        };
        
    \addplot[draw=none,fill=tuerkis] coordinates {
        (Battery,0)
        (Electrolyser,0)
        (PV,0) 
        (Hydrogen turbine,0)
        (Wind onshore,0) 
        (Wind offshore,0)
        (Network expansion, 0.8) 
      };
  \end{axis}

\end{tikzpicture}
	\caption{Comparison of installed capacities and network expansion between the \textit{REF} and the \textit{EFF} scenario (both with the integrated approach).}
	\label{fig:cap_plot2}
\end{figure}
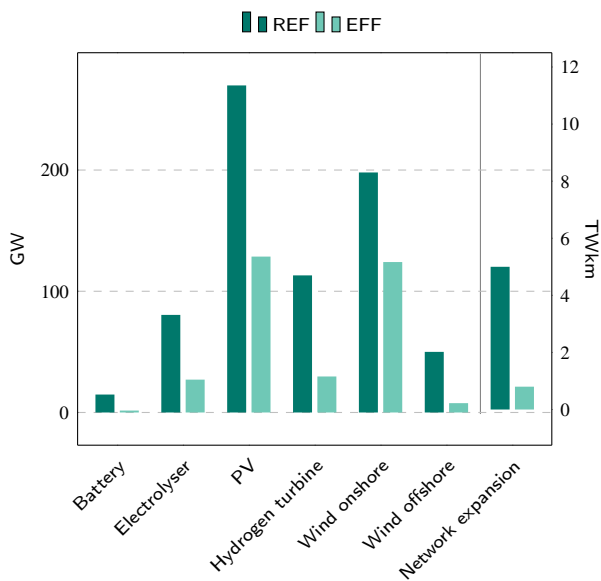

The consumption of resources, land, and materials caused by electricity generation can be reduced with systematic efficient measures. Fig. \ref{fig:cap_plot2} compares the result from using a reduced demand (\textit{EFF}) to the previous baseline assumption from \textit{REF}. However, this analysis does not take into account the costs of obtaining the efficiency gains proposed by the scenario. In both cases the integrated approach is used. A significant drop in the installed power of solar PV can be observed. This leads also to a reduced investment in batteries. The investments in wind onshore also decrease strongly and are almost equal to the installed capacity of solar PV. Wind offshore investments are further reduced  to 7 GW. The missing exogenous hydrogen demand and overall lower feed-in from renewables leads to  lower capacities in electrolysers since the surplus peaks are lower. The network expansion decreases to approximately one fifth since required peak capacity to transport electricity generated by wind from north to south is lower.

\subsection{Sufficient electricity generation in winter time}

\begin{figure*}
    \centering
    \includegraphics[width=.49\linewidth]{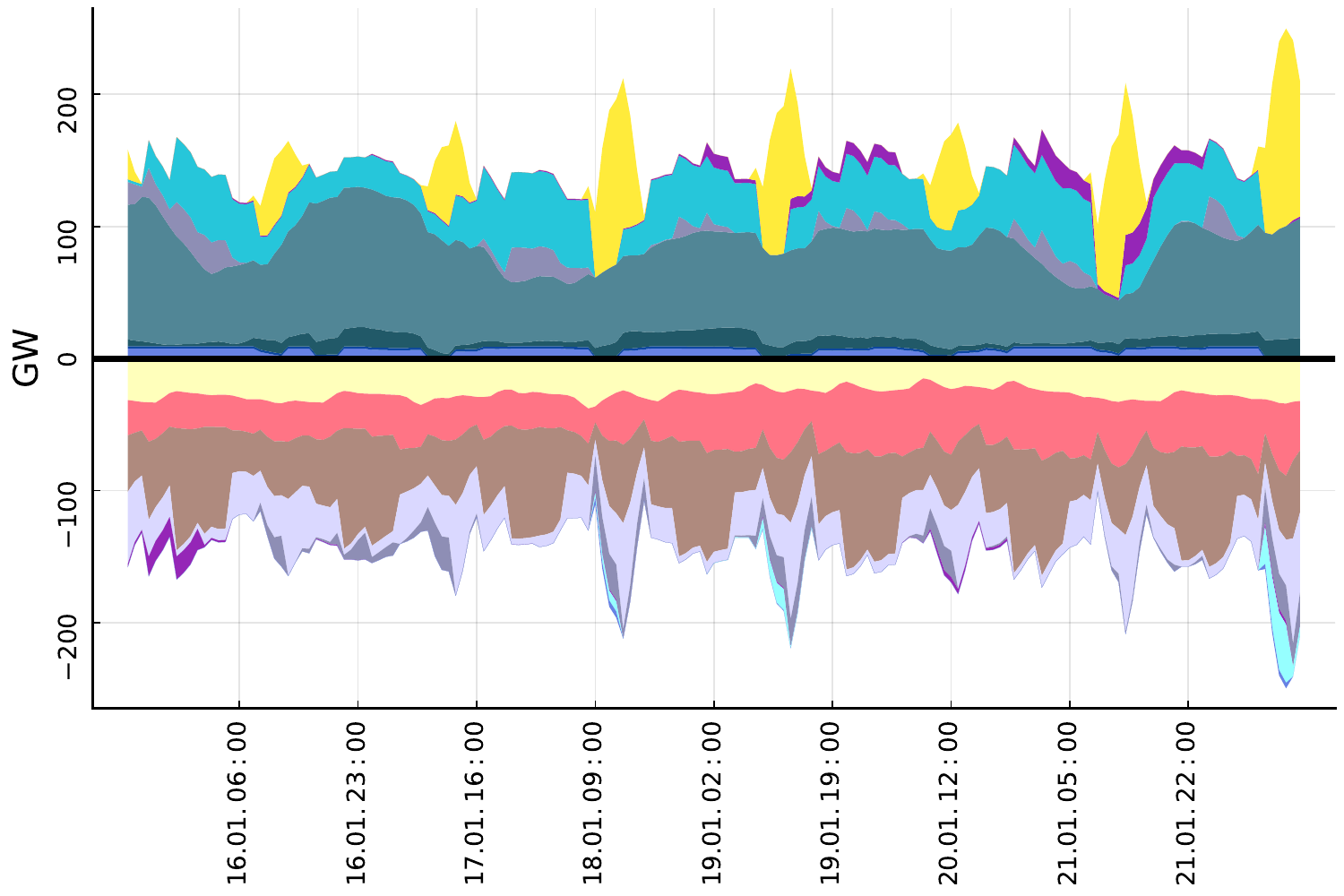}
    \includegraphics[width=.489\linewidth]{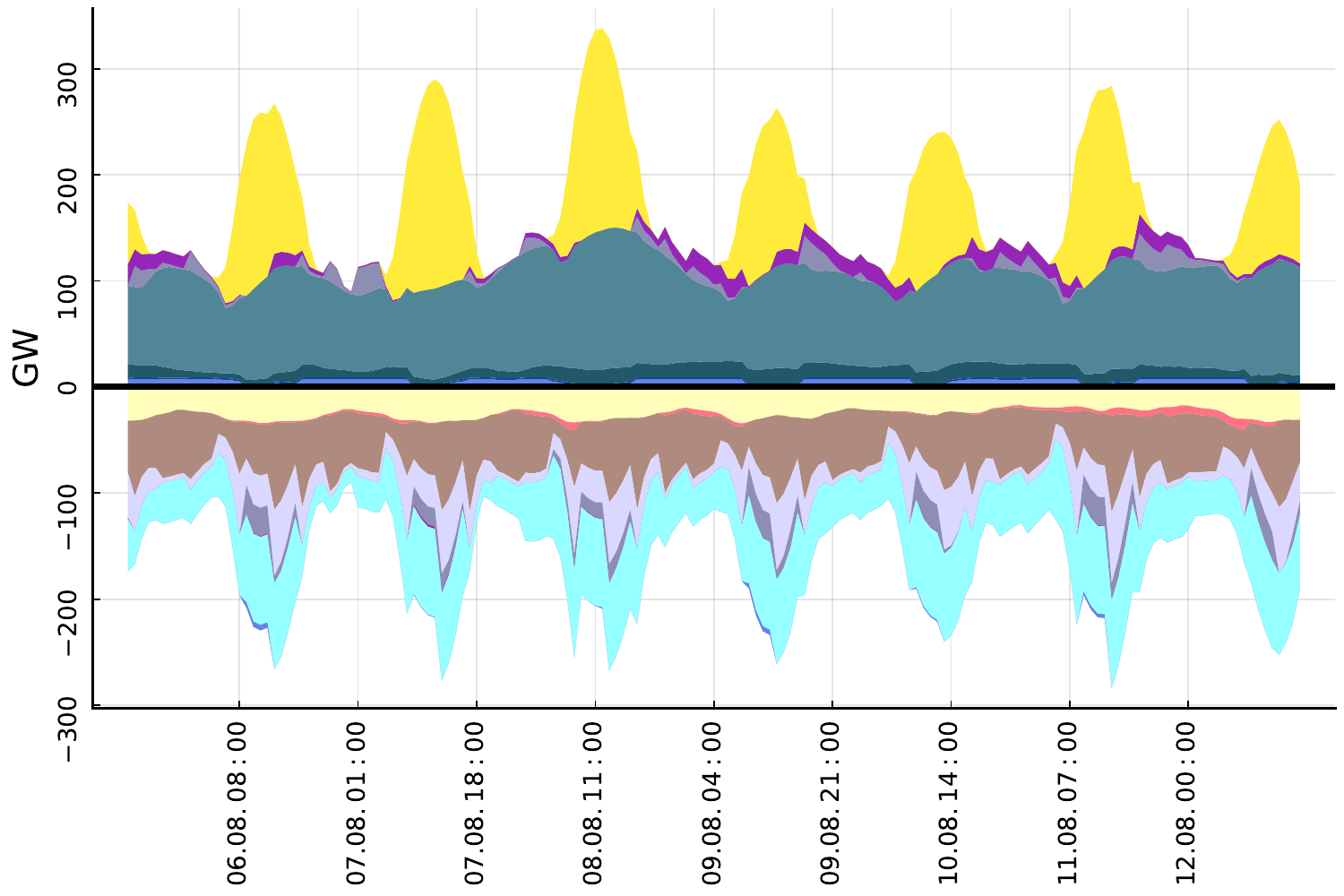}
    \includegraphics[width=.75\linewidth]{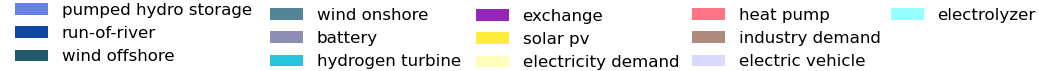}
	\caption{Exemplary generation and consumption of a winter week (left) and a summer week (right) in Germany for the integrated scenario.}
	\label{fig:dispatch}
\end{figure*}

Fig. \ref{fig:dispatch} depicts the electricity generation and consumption from the \textit{Integrated} scenario for a representative week in winter and summer. The generation in winter times consists of the feed-in from wind onshore with some solar PV peaks. The lacking electricity is generated by hydrogen turbines and some imports from European neighbours. On the other side, the consumption from heat pumps is quite high. Though they have some flexibility to shift their demand for a few hours, the consumption is rather constant at this state. In this situation the demand from industry and electric vehicles adapts to the generation pattern and contributes to the integration of renewable generation.

In the summer the picture changes. Besides a lower wind onshore generation, solar PV significantly increases the production which is characterised by the diurnal pattern. Compared to the winter week, there is almost a steady surplus of electricity which needs to be used. However, as the consumption from heat pumps is almost non existent, the demand is even lower than in the winter. Most of the excessive power is consumed by electrolysers and stored in form of hydrogen. This provides seasonal storage and also is needed in order to satisfy exogenous demand from hydrogen.

\section{Results II: Comparing decentral with central approaches}
\label{section4}

Although decentralized approaches and communal ownership prevailed at the beginning of electrification, today's energy system developed in a highly centralized manner in the 20th century. Generation sites were primarily resource-based (e.g. coal, hydroelectric) or policy-based (e.g. nuclear) and interconnected by long-range transmission grids. This approach, referred to as centralized in the following, leads to cost degression in power generation, but also to substantial grid expansion.

\subsection{Scenarios}

In the following, selected cost categories and sensitivities of the energy system are analyzed with respect to a decentralized approach with spatial proximity to consumption. In the decentralized scenario (\textit{Decentral}), the exchange between regions takes place via the already existing transmission grid. This scenario represents a world in which energy production is rather regional and local potentials and flexibility are used. Accordingly, the energy system tends to be operated in a more decentral way and specifics on site are taken into account to a greater extent. The central scenario (\textit{Central}) with 50 GW of offshore wind being built and unlimited grid expansion, represents the other extreme scenario.

To understand the gradual changes in between these two scenarios, we add two sensitiviy analysis. In the first one we set an upper limit for grid expansion in order to understand which grid corridors are prioritized. The second sensitivity analysis is performed for offshore wind, which is considered a central element in the comparison with photovoltaics and onshore wind. Due to the geography of Germany, the addition of offshore wind creates a north-south disparity, as energy demand tends to be located south. In order to find out how sensitive the rest of the system reacts to offshore wind, scenarios are calculated in which the expansion is step-wise specified in the range of 10 GW to 50 GW.

\subsection{The total cost are almost equal for the Central und Decentral scenario}

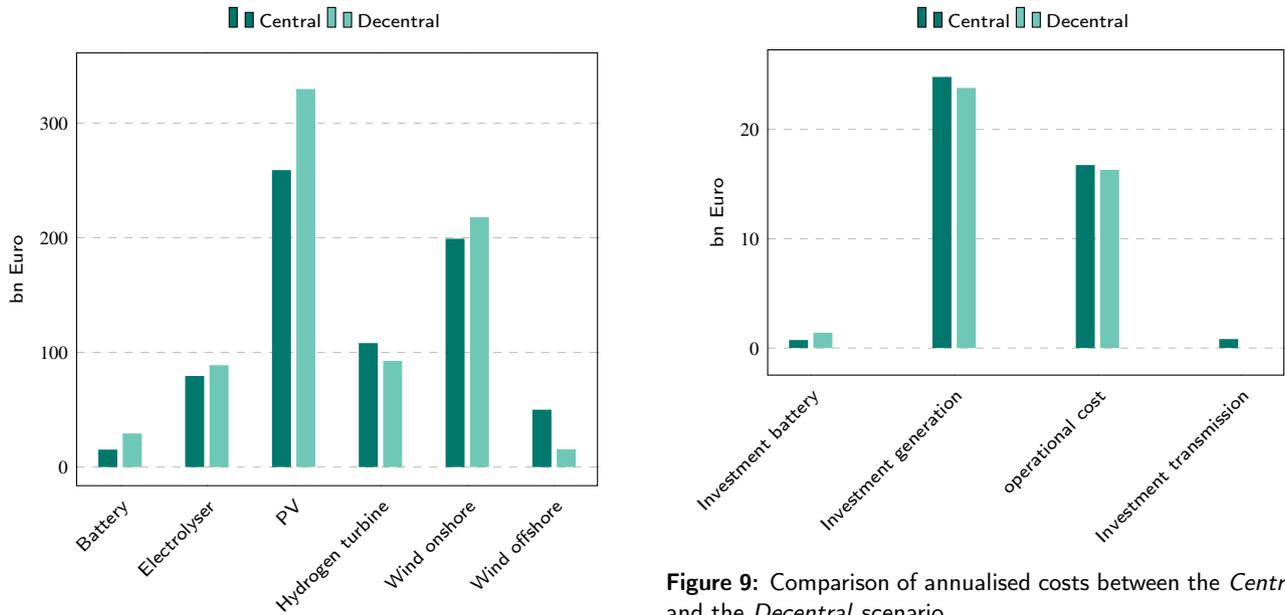
\begin{figure}
    \centering
\begin{tikzpicture}[font=\scriptsize]

  \begin{axis}[
    ybar,
    symbolic x coords={Battery, Electrolyser, PV, Hydrogen turbine, Wind onshore, Wind offshore},
    bar width=0.25cm,
    xtick=data,
    x tick label style={rotate=45,anchor=north east},
    ylabel=bn Euro,
    ylabel near ticks,
    ymajorgrids=true,
    grid style=dashed,
    tickwidth         = 0pt,
    legend style={
			draw=none,
			at={(0.5,1.12)},
			anchor=north},
	legend columns=-1
    ]
    
     \addplot[draw=none,fill=dunkelgruen] coordinates {
        (Battery,15.2)
        (Electrolyser,79.4)
        (PV,259.0)
        (Hydrogen turbine,108.1)
        (Wind onshore,199.1)
        (Wind offshore,50.0)
    };

    \addplot[draw=none,fill=tuerkis] coordinates {
        (Battery,29.3)
        (Electrolyser,88.8)
        (PV,329.7)
        (Hydrogen turbine,92.5)
        (Wind onshore,217.9)
        (Wind offshore,15.4)
      };
    
    \legend{Central, Decentral}
  \end{axis}
  
\end{tikzpicture}
	\caption{Comparison of installed capacities and network expansion between the \textit{Central} and the \textit{Decentral} scenario.}
	\label{fig:cap_plot3}
\end{figure}

As in section \ref{section3}, the biggest deviation between the two scenarios \textit{Central} and \textit{Decentral} is the installed capacity of solar PV and their respective deployment locations. The absence of network expansion in the \textit{Decentral} scenario results in a more regionalized pattern of generation and storage capacities. The wind onshore investment slightly increases to the maximal potential while wind offshore decreases from 50 GW to 15 GW (Fig. \ref{fig:cap_plot3}). The installed capacity of electrolysers increases a little due to higher solar PV investments and the associated stronger seasonality with higher feed-in peaks. Higher capacities of batteries contribute to cover the peak load which reduces the need for hydrogen turbines. A similar effect occurs for the deployment of solar PV and wind onshore turbines between the \textit{Central} and \textit{Decentral} scenarios. Most of the additional generation capacity is placed in the south-west and thus, closer to the regions with high electricity demand.

\begin{figure}
    \centering
\begin{tikzpicture}[font=\scriptsize]

  \begin{axis}[
    ybar,
    symbolic x coords={Investment battery, Investment generation, operational cost, Investment transmission},
    bar width=0.25cm,
    xtick=data,
    x tick label style={rotate=45,anchor=north east},
    ylabel=bn Euro,
    ylabel near ticks,
    ymajorgrids=true,
    grid style=dashed,
    tickwidth         = 0pt,
    legend style={
			draw=none,
			at={(0.5,1.15)},
			anchor=north},
	legend columns=-1,
	height=0.7\columnwidth,
	width=\columnwidth,
    ]
    
    \addplot[draw=none,fill=dunkelgruen] coordinates {
        (Investment battery,0.73)
        (Investment generation,24.8)
        (operational cost, 16.74)
        (Investment transmission,0.82)
    };
    
    \addplot[draw=none,fill=tuerkis] coordinates {
        (Investment battery,1.4)
        (Investment generation,23.8)
        (operational cost, 16.3)
        (Investment transmission,0)
      };
     
    \legend{Central, Decentral}
  \end{axis}
  
\end{tikzpicture}
	\caption{Comparison of annualised costs between the \textit{Central} and the \textit{Decentral} scenario.}
	\label{fig:cost_plot}
\end{figure}

The total cost of both scenarios are almost the same and differ less than 0.5\% of the total cost. Considering the amount of uncertain parameters in the model that had to be assumed and estimated, this difference is negligeable. The higher expenditures in batteries in the \textit{Decentral} scenario are offset by lower investments in the generation capacites, operational costs, and network expansion investments (Fig. \ref{fig:cost_plot}).

\subsection{Network expansion can be substituted using PV with storage options}

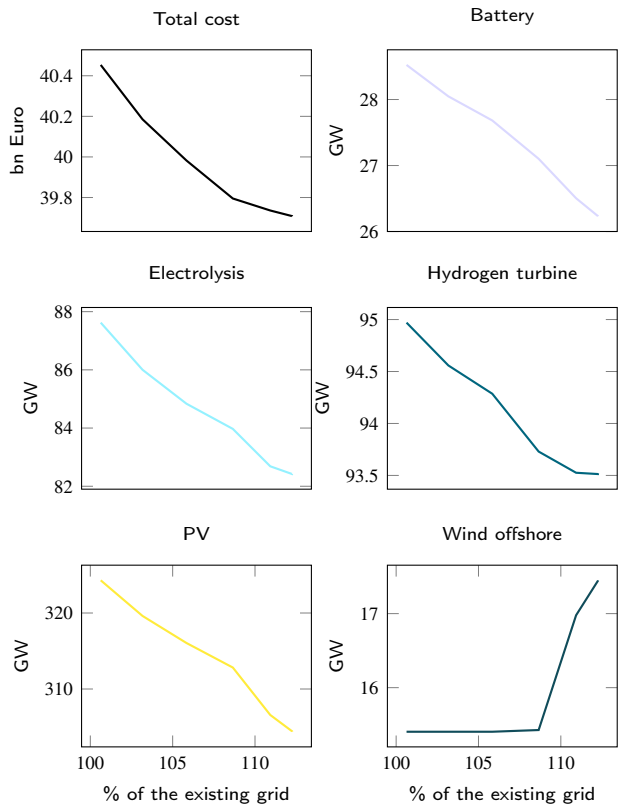
\begin{figure}
    \centering
\begin{tikzpicture}[font=\scriptsize]
  \begin{groupplot}[
    group style={group size=2 by 3},
    width=0.55\columnwidth,
    no markers,
    ylabel near ticks,
    xlabel near ticks
    ]
    
    \nextgroupplot[title=Total cost, ylabel=bn Euro, xmajorticks=false]
        \addplot[thick, color=black] table [x=scen, y=Gesamtkosten, col sep=comma]  {figs/data/sensitivity_ntc_data.csv};

    \nextgroupplot[title=Battery, ylabel=GW, xmajorticks=false]
        \addplot[thick,color=batterycolor] table [x=scen, y=battery, col sep=comma]  {figs/data/sensitivity_ntc_data.csv};

    \nextgroupplot[title=Electrolysis, ylabel=GW, xmajorticks=false]
        \addplot[thick,color=electrolysiscolor] table [x=scen, y=electrolysis, col sep=comma]  {figs/data/sensitivity_ntc_data.csv};
 
    \nextgroupplot[title=Hydrogen turbine, ylabel=GW, xmajorticks=false]
        \addplot[thick,color=hydrogencolor] table [x=scen, y=ocgtHydrogen, col sep=comma]  {figs/data/sensitivity_ntc_data.csv};
    
    \nextgroupplot[title=PV, ylabel=GW, xlabel=\% of the existing grid]
        \addplot[thick,color=pvcolor] table [x=scen, y=pv, col sep=comma]  {figs/data/sensitivity_ntc_data.csv};
	  
	 \nextgroupplot[title=Wind offshore, ylabel=GW, xlabel=\% of the existing grid]
        \addplot[thick,color=windoffcolor] table [x=scen, y=wind_offshore, col sep=comma]  {figs/data/sensitivity_ntc_data.csv};
        
  \end{groupplot}
\end{tikzpicture}

	\caption{Installed capacities subject to a upper bound of network expansion (100\% corresponds to today's grid).}
	\label{fig:sensitivity_grid}

\end{figure}

We add two senstivity analyses, starting with a focus on network expansion: From the existing grid with the assumption of no expansion, the possible network extension is raised gradually. Figure \ref{fig:sensitivity_grid} shows the results of the gradual expansion of the grid infrastructure as a substition curve for various technologies. The x-axis shows the network expansions in relation the existing grid in percent. As to be expected from previous results, solar PV has the strongest negative correlation. The required electrolyzers and battery storages are slightly reduced by more grid expansion, although not to the same extent as solar PV. There was hardly any increase of offshore wind in the span studied, which is due to the fact that it would require significantly higher amounts of grid expansion. Additional wind offshore capacity becomes valuable when certain threshold of grid expansion is reached which allows the produced electricity to be transported. Besides wind offshore the other technologies either have a linear correlation or do not change at all. The total costs consequently also decrease in a linear way.

\subsection{Wind offshore requires more network expansion}

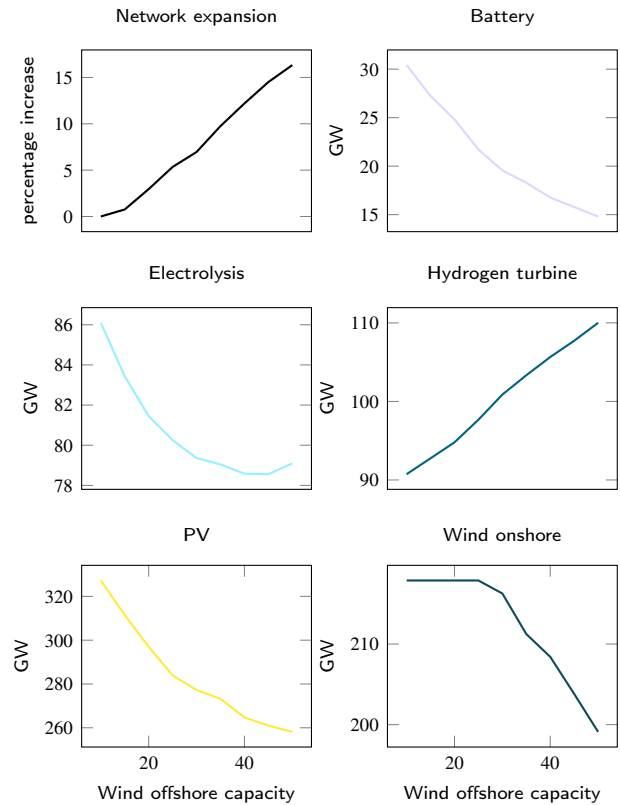
\begin{figure}
    \centering
\begin{tikzpicture}[font=\scriptsize]
  \begin{groupplot}[
    group style={group size=2 by 3},
    width=0.55\columnwidth,
    no markers,
    ylabel near ticks,
    xlabel near ticks
    ]
    
    \nextgroupplot[title=Network expansion, ylabel=percentage increase, xmajorticks=false]
        \addplot[thick, color=black] table [x=scen, y=Netzausbau, col sep=comma]  {figs/data/installed_capacites_ntc_offshore_data.csv};

    \nextgroupplot[title=Battery, ylabel=GW, xmajorticks=false]
        \addplot[thick,color=batterycolor] table [x=scen, y=battery, col sep=comma]  {figs/data/installed_capacites_ntc_offshore_data.csv};

    \nextgroupplot[title=Electrolysis, ylabel=GW, xmajorticks=false]
        \addplot[thick,color=electrolysiscolor] table [x=scen, y=electrolysis, col sep=comma]  {figs/data/installed_capacites_ntc_offshore_data.csv};
 
    \nextgroupplot[title=Hydrogen turbine, ylabel=GW, xmajorticks=false]
        \addplot[thick,color=hydrogencolor] table [x=scen, y=ocgtHydrogen, col sep=comma]  {figs/data/installed_capacites_ntc_offshore_data.csv};
    
    \nextgroupplot[title=PV, ylabel=GW, xlabel=Wind offshore capacity]
        \addplot[thick,color=pvcolor] table [x=scen, y=pv, col sep=comma]  {figs/data/installed_capacites_ntc_offshore_data.csv};
	  
	 \nextgroupplot[title=Wind onshore, ylabel=GW, xlabel=Wind offshore capacity]
        \addplot[thick,color=windoffcolor] table [x=scen, y=wind_onshore, col sep=comma]  {figs/data/installed_capacites_ntc_offshore_data.csv};
        
  \end{groupplot}
\end{tikzpicture}
	\caption{Installed capacities subject to a fixed installed capacity of wind offshore in GW.}
	\label{fig:sensitivity_offshore}
\end{figure}

With the same method a sensitivity analysis was performed on the installed wind offshore capacity. Starting from 10 GW it was step-wise increased to 50 GW. Fig. \ref{fig:sensitivity_offshore} depicts the change of other parameters based on the exogenously set wind offshore capacity. The investments in solar PV and batteries are reduced and decline quite constantly. The wind onshore capacity is reduced if the wind offshore capacity is higher than 25 GW wind offshore capacity. Also, electrolyser capacity declines for the first 30 GW of wind offshore capacity but then change very litte towards 50 GW. Contrary to the installed power of batteries, the installed capacity of hydrogen turbines increases by the same amount in order to cover the peak load. The additional necessary network expansion grows linearly to the wind offshore capacity by approximately 4\% per 10 GW.

\FloatBarrier
\section{Conclusion and Policy Implications}
\label{section6}

To the best of our knowledge, this is the first paper to discuss a fully renewable energy system in the whole European Union, in Germany and on the level of all 38 NUTS-2 regions. We are interested in the trade-of between centralized grid expansion and a more decentral structure of electricity generation and storage, and the spatial variation within the German energy system, always considered in the context of the European energy system. As the energiewende is proceeding, additional information on spatial and temporal generation and demand patterns are required. This paper uses a very detailed energy system model to shed light on these questions, with a focus on different scenarios implemented for Germany. To keep modelling and results trackable, we do not explore similar scenarios, or different values for renewable potentials at the broader European level.

Amongst other things, we find that the consideration of grid expansion costs leads to a significant decrease of grid expansion and strengthens decentral generation closer to loads. An hourly resolution of the results shows that security of supply is guaranteed in all regions, even in cold winter weeks. Grid expansion represents one of several flexibility options, but in the status quo, neglecting grid costs leads to an excessive strengthening of centralized power supply, especially with offshore wind. Decentralized approaches with spatial proximity to consumption do not show significant cost differences compared to centralized approaches. The slightly higher investments in generation and battery storage in the decentralized scenario are offset by lower grid expansion costs. We do not go into detail with respect to other types of costs (such as land consumption and external costs of grid expansion, resource use for batteries and other equipment, etc., but conclude that there is no prior for the central neither for the decentral solution. Energy savings and efficiency are key drivers on the way to reduced generation structures and resource consumption.

The paper sheds light on the classical debate about the localization of generation and demand in the context of decarbonization. We can draw two types of conclusions, one focusing on modelling issues, and one focusing on policy issues. With respect to modelling issues, we find that a comparison between different institutional arrangements of calculating optimal investments and dispatch is important. Clearly, the two idealtypes that we have considered yield different results. From a welfare (or cost minimization) perspective, clearly the costs of network transmission should be taken into account in the overall optimization algorithm. Neglecting these costs may distort the allocation of electricity generating capacities. Our results also highlight the importance of taking into account two “new” elements of the system: electrolyzers, and storage infrastructure.

A concrete policy implication is that the current network ordnance in Germany should be changed, to reap the potential benefits from the overall cost minimization. Given the importance of flexibility, at least a part of storage capacity should be considered as part of the (regulated) infrastructure, which would facilitate investments. At a more general level, our paper suggests the need to combine national energy system planning with the European process, at least with the European neighbours.
Last but not least, the paper also indicates a potential to save grid infrastructure and generation assets for the "energy saving-efficiency" scenario with reduced energy demand. Even though the precise costs for the efficiency measures were not quantified, it seems to make sense to strengthen incentives for efficient energy use.

\section*{Acknowledgements}
The paper is a follow-up of Professor Claudia Kemfert’s keynote presentation at the IX. Academic Symposium in Bar\-ce\-lona entitled “Chance for decentralized energy system transformation with full supply from renewable energies”; it is based on a larger research program on the German energiewende. The scenarios and results are also based on a cooperation with the 100 prozent erneuerbar foundation (Berlin). We thank the organizers and participants at the Barcelona symposium in particular Maria Teresa Costa-Campi and Elisa Trujillo-Baute, two anonymous referees for particularly useful and detailed comments, furthermore participants at the DIW brown bag (März 2021), the “Berlin Energy Days” (April 2021), and others colleagues for comments and suggestions. The usual disclaimer applies, i.e. the authors alone are responsbile for the content.

\bibliography{sources}
\bibliographystyle{cas-model2-names}

\end{document}